\documentclass[10pt,twocolumn]{IEEEtran}
\usepackage{cite}
\usepackage{amssymb}
\usepackage{amsmath}
\usepackage{dblfloatfix}
\usepackage{amsthm}
\usepackage{graphicx}
\graphicspath{{./graphics/}{.}}
\usepackage{color}
\usepackage{multirow}

\newcommand{\hd}{\hdots}
\newcommand{\hs}[1]{\hspace{-#1 ex}}
\long\def\symbolfootnote[#1]#2{\begingroup
\def\thefootnote{\fnsymbol{footnote}}\footnote[#1]{#2}\endgroup}
\newcommand{\Exp}[1]{\mathbb{E}\left\{#1\right\}}

\newcommand{\bs}[1]{\boldsymbol{ #1}}

\IEEEoverridecommandlockouts

\begin{document}
\title {EM-based Semi-blind Channel Estimation in AF Two-Way Relay Networks}
\author{Saeed Abdallah and Ioannis N. Psaromiligkos\thanks{The authors are with the Electrical and Computer Engineering Dept., McGill University, Montreal, Quebec, Canada. Email: saeed.abdallah@mail.mcgill.ca; yannis@ece.mcgill.ca.}\\
}

\maketitle
\begin{abstract}
We propose an expectation maximization (EM)-based algorithm for semi-blind channel estimation of reciprocal channels in amplify-and-forward (AF) two-way relay networks (TWRNs). By incorporating both data samples and pilots into the estimation, the proposed algorithm provides substantially higher accuracy than the conventional training-based approach. Furthermore, the proposed algorithm has a linear computational complexity per iteration and converges after a small number of iterations. 
\end{abstract}

\section{Introduction}
Amplify-and-forward (AF) two-way relay networks (TWRNs)~\cite{rankov07} require highly accurate information about the channel at the terminals, for both self-interference cancellation and coherent decoding. The problem of channel estimation for AF TWRNs has been considered in a number of works~\cite{gao2009optimal,gao2009ofdm,feifei_power,tensor_based}. Most works adopt a training-based approach that requires each terminal to transmit a pilot sequence known to the other terminal. Despite the robustness and simplicity of this approach, the transmission of known pilots consumes bandwidth resources, which undermines the spectral efficiency of TWRNs. 

One alternative to training-based estimation is semi-blind estimation~\cite{semiblind3} which, in addition to using pilots, also incorporates the received data samples into the estimation. Semi-blind estimation requires a smaller number of pilots, which makes it more spectrally efficient than training-based estimation~\cite{semiblind3}. 

In this paper we derive a semi-blind expectation-maximization (EM)-based channel estimator for AF TRWNs, assuming reciprocal flat-fading channels.  Each iteration of the proposed algorithm has a computational complexity that is linear in the number of data samples. Using simulations, we show that, even with a limited number of data samples, the EM algorithm provides substantially better accuracy than the training-based least-squares (LS) estimator. Only a small number of iterations is needed to achieve convergence. 

The rest of this work is organized as follows. In Section~\ref{system_model_ch6} we present the system model. In Section~\ref{proposed_ch6}, we derive the EM algorithm. Simulation results are presented in Section~\ref{simulations_ch6}. Finally, our conclusions are discussed in Section~\ref{conclusions_ch6}.

\section{System Model}
\label{system_model_ch6}




We consider a half-duplex AF TWRN with two terminals $\mathcal{T}_1$, $\mathcal{T}_2$ and a single relay $\mathcal{R}$ operating in flat-fading channel conditions. Each transmission period is divided into two phases:

\noindent \textbf{Phase 1}: Each terminal transmits a block of $L$ pilot symbols, followed by $N$ data symbols. We denote by $\bs{t}_1\triangleq[t_{11},\hd,t_{1L}]^{T}$ and $\bs{t}_2\triangleq[t_{21},\hd,t_{2L}]^{T}$ the pilot symbol vectors of $\mathcal{T}_1$ and $\mathcal{T}_2$, and by $\bs{s}_1\triangleq[s_{11},\hd,s_{1N}]^{T}$ and $\bs{s}_2\triangleq[s_{21},\hd,s_{2N}]^{T}$ the transmitted data symbol vectors of $\mathcal{T}_1$ and $\mathcal{T}_2$, respectively. The data symbols $s_{21},\hd,s_{2N}$ are equiprobably drawn from the set $S=\{\xi_1,\hd,\xi_M\}$ of size $M$. The average transmission powers of $\mathcal{T}_1$ and $\mathcal{T}_2$ during pilot transmission are $P_1$ and $P_2$, respectively, i.e., $\bs{t}_1^{H}\bs{t}_1=LP_1$ and $\bs{t}_2^{H}\bs{t}_2=LP_2$. For simplicity, we assume that each terminal uses the same power during pilot and data transmission, i.e., $\Exp{\bs{s}_1^{H}\bs{s}_1}=NP_1$ and $\Exp{\bs{s}_2^{H}\bs{s}_2}=NP_2$. The corresponding received pilot and data signal vectors at $\mathcal{R}$ are $\bs{r}_P=h\bs{t}_1+g\bs{t}_2+\bs{\omega}$, and $\bs{r}_D=h\bs{s}_1+g\bs{s}_2+\bs{n}$ where $\bs{\omega}$ and $\bs{n}$ are additive white Gaussian noise vectors with mean zero and covariance matrices\footnote{$\bs{I}_L$ denotes the $L\times L$ identity matrix.} $\sigma^2\bs{I}_L$ and $\sigma^2\bs{I}_N$, respectively. 

\noindent \textbf{Phase 2}: The relay broadcasts the vectors $A\bs{r}_P$ and $A\bs{r}_D$ in sequence, where $A>0$ is the amplification factor. The corresponding received pilot signal vector at $\mathcal{T}_1$ is 
\begin{equation}
\bs{y}=Ah^2\bs{t}_1+Ahg\bs{t}_2+Ah\bs{\omega}+\bs{\omega}_1
\end{equation}
and the received data signal vector is
\begin{equation}
\bs{z}=Ah^2\bs{s}_1+Ahg\bs{s}_2+Ah\bs{n}+\bs{n}_1
\end{equation}
where $\bs{\omega}_1$ and $\bs{n}_1$ are also additive white Gaussian noise vectors with mean zero and covariance matrices $\sigma^2\bs{I}_L$ and $\sigma^2\bs{I}_N$, respectively. The average transmission power of the relay is maintained at $P_r$ over the long term by using the amplification factor $A=\sqrt{\frac{P_r}{P_1+P_2+\sigma^2}}$. 
We are interested in the estimation of the cascaded channel parameters $a\triangleq h^2$ and $b\triangleq hg$, which are sufficient for detection. The channel coefficients remain fixed during the transmission of the $L$ pilots and $N$ data symbols. Moreover, the noise variance $\sigma^2$ is assumed to be known at $\mathcal{T}_1$.

\section{Proposed Channel Estimation Algorithm}
\label{proposed_ch6}
Let $\bs{\theta}\triangleq[a,b]^{T}$ be the vector of unknown parameters that we wish to estimate. The observed vectors $\{\bs{y},\bs{z}\}$ represent the incomplete data and the data symbols $\bs{s}_2$ represent the hidden data. Hence, the complete data is $\{\bs{y},\bs{z},\bs{s}_2\}$, and the corresponding log-likelihood function (LLF) is 
\begin{equation}
\label{loglikelihood_reciprocal}
\begin{split}
&\mathcal{L}(\bs{y},\bs{z},\bs{s}_2;\bs{\theta})\hs{0.3}=\hs{0.3}-\hs{0.3}N\log M\hs{0.5}-\hs{0.5}(N\hs{0.3}+\hs{0.3}L)\log(\pi \sigma^2(A^2|a|+1))-\\&\frac{1}{\sigma^2(A^2|a|+1)}\hs{0.3}\left(\hs{0.3}\big\Vert\bs{y}\hs{0.5}-\hs{0.5}Aa\bs{t}_1\hs{0.5}+\hs{0.5}Ab\bs{t}_2\big\Vert^2
\hs{0.3}+\hs{0.3}\big\Vert\bs{z}-Aa\bs{s}_1-Ab\bs{s}_2\big\Vert^2\hs{0.3}\right)\hs{0.4}.
\end{split}
\end{equation}

An iteration of the EM algorithm, say the $t$th one, consists of two steps. The first step, called the expectation step (E-step) consists of evaluating the expectation
\begin{equation}
Q(\bs{\theta};\bs{\theta}^{(t)})=\Exp{\mathcal{L}(\bs{y},\bs{z},\bs{s}_2;\bs{\theta})|\bs{y},\bs{z};\bs{\theta}^{(t)}}
\end{equation} 
of the LLF of the complete data, $\mathcal{L}(\bs{y},\bs{z},\bs{s}_2;\bs{\theta})$, with respect to the conditional PMF $f(\bs{s}_2|\bs{y},\bs{z};\bs{\theta}^{(t)})$ of the hidden data given the observations and the current estimate $\bs{\theta}^{(t)}\triangleq[a^{(t)},b^{(t)}]^{T}$ of $\bs{\theta}$. The second step of the EM algorithm is the maximization step (M-step) which consists of maximizing the expectation $Q(\bs{\theta};\bs{\theta}^{(t)})$ with respect to $\bs{\theta}$ to obtain an updated estimate $\bs{\theta}^{(t+1)}$, i.e.,
\begin{equation}
\bs{\theta}^{(t+1)}=\arg\max_{\bs{\theta}}Q(\bs{\theta};\bs{\theta}^{(t)}).
\end{equation}
In our case, the E-step and M-step are as follows:

\noindent\textbf{E-step}: We have\footnote{We ignore the term $N\log M$ since it has no impact on the solution.}
\begin{equation}
\label{Estep_reciprocal}
\begin{aligned}
&Q\big(\bs{\theta};\bs{\theta}^{(t)}\big)\hs{0.4}=\hs{0.4}-(N\hs{0.4}+\hs{0.4}L)\log(\pi \sigma^2(A^2|a|\hs{0.4}+\hs{0.4}1))\hs{0.4}-\hs{0.4}\frac{1}{\sigma^2(A^2|a|+1)}\hs{0.3}\times\\&\hs{0.3}\left(\hs{0.3}\Vert\bs{y}\hs{0.3}-\hs{0.3}Aa\bs{t}_1\hs{0.3}-\hs{0.3}Ab\bs{t}_2\Vert^2
\hs{0.3}+\hs{0.3}\sum\limits_{i=1}^{N}\sum\limits_{j=1}^{M}\beta_{i,j}^{(t)}\big|z_i\hs{0.3}-\hs{0.3}Aat_{1i}\hs{0.3}-\hs{0.3}Ab\xi_j\big|^2\hs{0.3}\right)\hs{0.5}
\end{aligned}
\end{equation}
where $\beta_{i,j}^{(t)}$ is the posterior PMF of the $i$th data symbol during the $t$th iteration, given by
\begin{equation}
\begin{split}
\beta_{i,j}^{(t)}=\frac{e^{-\frac{1}{\sigma^2(A^2|a^{(t)}|+1)}|z_i-Aa^{(t)}s_{1i}-Ab^{(t)}\xi_j|^2}}{\sum\limits_{k=1}^{M}e^{-\frac{1}{\sigma^2(A^2|a^{(t)}|+1)}|z_i-Aa^{(t)}s_{1i}-Ab^{(t)}\xi_k|^2}}.
\end{split}
\end{equation}

\noindent\textbf{M-step}: We need to obtain the values $a^{(t+1)}$, $b^{(t+1)}$ such that 
\begin{equation}
\label{Mstep_reciprocal}
\big\{a^{(t+1)},b^{(t+1)}\big\}=\arg\max\limits_{\bs{\theta}=[a,b]^{T}} Q\big(\bs{\theta};\bs{\theta}^{(t)}\big).
\end{equation}
Regarding $b^{(t+1)}$, it can be easily verified that the value of $b$ that maximizes $Q(\bs{\theta};\bs{\theta}^{(t)})$ for a given value of $a$ is 
\begin{equation}
\label{boa}
b_o(a)=\frac{\sum\limits_{i=1}^{N}\sum\limits_{j=1}^{M}\beta_{i,j}^{(t)}\xi_j^{*}\left(z_i-Aas_{1i}\right)+\bs{t}_2^{H}\bs{y}-Aa\bs{t}_2^{H}\bs{t}_2}{A\sum\limits_{i=1}^{N}\sum\limits_{j=1}^{M}\beta_{i,j}^{(t)}+A\bs{t}_2^{H}\bs{t}_2}.
\end{equation}
Substituting $b_o(a)$ in place of $b$ in~\eqref{Estep_reciprocal}, we obtain an updated objective function that depends only on $a$. This function, which we denote as $Q(a;\bs{\theta}^{(t)})$, is given by
\begin{equation}
\label{updated1}
\begin{aligned}
&Q(a;\bs{\theta}^{(t)})=-N\log(\pi\sigma^2(A^2|a|+1))\hs{0.3}-\hs{0.3}\frac{1}{G^2\sigma^2(A^2|a|+1)}\times\\&\Bigg(\big\Vert G\bs{y}\hs{0.3}-\hs{0.3}AaG\bs{t}_1\hs{0.3}-\hs{0.3}A\mathcal{I}\bs{t}_2\hs{0.3}+\hs{0.3}A^2a\mathcal{X}\bs{t}_2\big\Vert^2+\\
&\sum\limits_{i=1}^{N}\sum\limits_{j=1}^{M}\beta_{i,j}^{(t)}\big|Gz_i-AaGs_{1i}-A\mathcal{I}\xi_j+A^2a\mathcal{X}\xi_j\big|^2\Bigg)
\end{aligned}
\end{equation}
where
\begin{equation*}
G\hs{0.3}=\hs{0.3}A\hs{0.5}\left(\hs{0.3}\sum\limits_{i=1}^{N}\sum\limits_{j=1}^{M}\hs{0.3}\beta_{i,j}^{(t)}|\xi_j|^2\hs{0.3}+\hs{0.3}\bs{t}_2^{H}\bs{t}_2\hs{0.3}\right)\hs{0.5},\ \mathcal{I}\hs{0.3}=\hs{0.3}\sum\limits_{i=1}^{N}\sum\limits_{j=1}^{M}\hs{0.4}\beta_{i,j}^{(t)}\xi_j^{*}\bs{z}_i+\bs{t}_2^{H}\bs{y},
\end{equation*}
\begin{equation*} \mbox{and}\ \ \ \ \ \ \ \ \ \mathcal{X}=\sum\limits_{i=1}^{N}\sum\limits_{j=1}^{M}\beta_{i,j}^{(t)}\xi_j^{*}s_{1i}+\bs{t}_2^{H}\bs{t}_1.
\end{equation*}
In order to maximize~\eqref{updated1} with respect to $a$, we will maximize it first with respect to the phase $\phi_a\triangleq\angle a$ and then with respect to $|a|$. 
Minimizing~\eqref{updated1} with respect to $\phi_a$, we obtain
\begin{equation}
\label{phia_rec}
\begin{aligned}
&\phi_a^{(t+1)}\hs{0.3}=\hs{0.3}\pi\hs{0.3}-\hs{0.3}\angle\bigg(\hs{0.3}\sum\limits_{i=1}^{N}\sum\limits_{j=1}^{M}\beta_{i,j}^{(t)}
\big(Gz_i\hs{0.3}-\hs{0.3}A\mathcal{I}\xi_j\big)^{*}\big(A^2\mathcal{X}\xi_j\hs{0.3}-\hs{0.3}AGs_{1i}\big)\hs{0.3}\\
&\ \ \ +\big(G\bs{y}\hs{0.3}-\hs{0.3}A\mathcal{I}\bs{t}_2\big)^{H}
\big(A^2\mathcal{X}\bs{t}_2\hs{0.3}-\hs{0.3}AG\bs{t}_1\big)\bigg).
\end{aligned}
\end{equation}
Substituting $\phi_a^{(t+1)}$ into~\eqref{updated1}, we get the following function that depends only on $|a|$:
\begin{equation}
\label{updated2}
Q(|a|;\bs{\theta}^{(t)})\hs{0.3}=\hs{0.3}-N\log(\pi\sigma^2(A^2|a|+1))-\frac{\breve{V}|a|^2\hs{0.3}-\hs{0.3}2\breve{W}|a|\hs{0.3}+\hs{0.3}\breve{U}}{\sigma^2(A^2|a|+1)},
\end{equation}
where
\begin{equation*}
\breve{U}=\frac{1}{G^2}\sum\limits_{i=1}^{N}\sum\limits_{j=1}^{M}\beta_{i,j}^{(t)}\big|Gz_i-A\mathcal{I}\xi_j\big|^2+\frac{1}{G^2}\big\Vert G\bs{y}-A\mathcal{I}\bs{t}_2\big\Vert^2,
\end{equation*}
\begin{equation*}
\breve{V}\hs{0.4}=\hs{0.4}\frac{1}{G^2}\hs{0.2}\sum\limits_{i=1}^{N}\hs{0.2}\sum\limits_{j=1}^{M}\hs{0.4}\beta_{i,j}^{(t)}\big|A^2\mathcal{X}\xi_j-AGs_{1i}\big|^2\hs{0.2}+\frac{1}{G^2}\big\Vert A^2\mathcal{X}\bs{t}_2-AG\bs{t}_1\big\Vert^2
\end{equation*}
\begin{equation*}
\begin{split}
&\breve{W}=\frac{1}{G^2}\bigg|\sum\limits_{i=1}^{N}\sum\limits_{j=1}^{M}\beta_{i,j}^{(t)}\big(Gz_i-A\mathcal{I}\xi_j\big)^{*}\big(A^2\mathcal{X}\xi_j-AGs_{1i}\big)+\\&\ \ \ \big(G\bs{y}-A\mathcal{I}\bs{t}_2\big)^{H}\big(A^2\mathcal{X}\bs{t}_2-AG\bs{t}_1\big)\bigg|.
\end{split}
\end{equation*}
Taking the derivative of~\eqref{updated2} w.r.t. $|a|$, we get
\begin{equation}
\frac{dQ(|a|;\bs{\theta}^{(t)})}{d|a|}\hs{0.3}=\hs{0.2}-\hs{0.1}\frac{NA^2}{A^2|a|+1}\hs{0.1}-\hs{0.1}\frac{A^2\breve{V}|a|^2\hs{0.3}+\hs{0.3}2\breve{V}|a|\hs{0.3}-\hs{0.3}A^2\breve{U}\hs{0.3}-\hs{0.3}2\breve{W}}{\sigma^2(A^2|a|+1)^2}.
\end{equation}
Setting $\frac{dQ(|a|;\bs{\theta}^{(t)})}{d|a|}=0$, we then obtain the quadratic equation
\begin{equation}
\label{quadratic}
A^2\breve{V}|a|^2+(2\breve{V}+NA^4\sigma^2)|a|+NA^2\sigma^2-A^2\breve{U}-2\breve{W}=0.
\end{equation}
Solving~\eqref{quadratic}, we finally get
\begin{equation}
\label{xi_rec}
\begin{split}
&|a|^{(t+1)}=\frac{1}{2A^2\breve{V}}\bigg(-(2\breve{V}+NA^4\sigma^2)+\\& \sqrt{(2\breve{V}+NA^4\sigma^2)^2-4A^2\breve{V}(NA^2\sigma^2-A^2\breve{U}-2\breve{W})}\bigg).
\end{split}
\end{equation}
As we can see from~\eqref{boa},~\eqref{phia_rec} and~\eqref{xi_rec}, the computational complexity of each EM iteration is $O(N)$, i.e., it is linear in the number of data samples.


\section{Simulation Results}
\label{simulations_ch6}

In this section, we investigate through simulations the MSE performance of the derived EM algorithm. We model $h$ and $g$ as independent complex Gaussian RVs with mean zero and variance 1. We assume that $P_1=P_2=P_r$ and average our results over $100$ independent channel realizations. The data symbols are generated from square QAM constellations. We consider the modulation orders $M=4,\ 16,\ 64$. The pilot vectors $\bs{t}_1$ and $\bs{t}_2$ are obtained using $M=4$ and chosen to be orthogonal. In our plots, we consider the total MSE, which is the sum of the MSE for the estimation of $a$ and $b$. 

\begin{figure}[!ht]
\centering
\includegraphics[width=3.4in, height=2.5in]{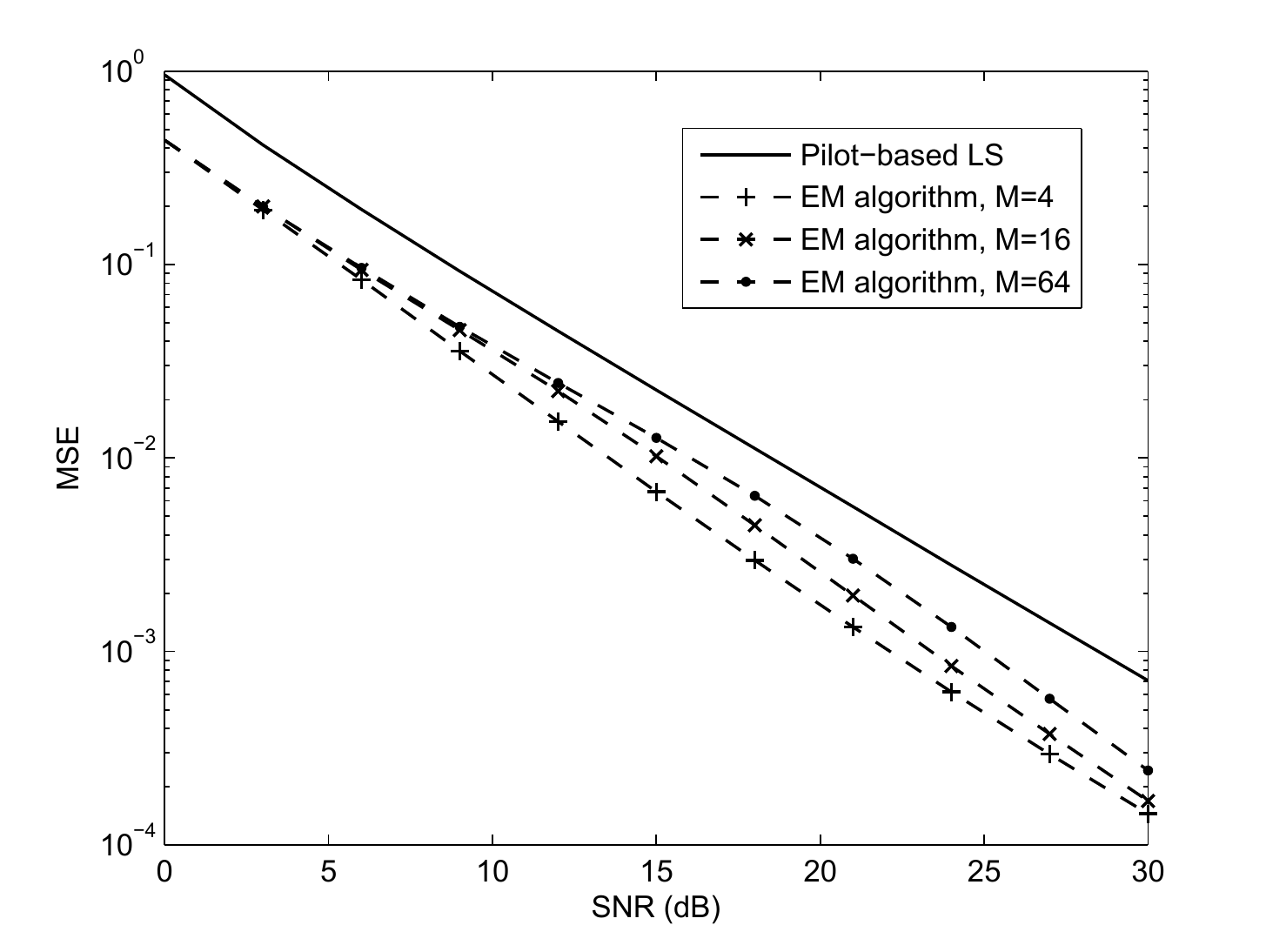}
\caption[MSE of EM algorithm vs. SNR (reciprocal channels, $4$ EM iterations).]{MSE performance of the EM algorithm for reciprocal channels along with the corresponding semi-blind CRBs plotted versus SNR ($N=32$ and $L=8$, 4 EM iterations).}
\label{MSE_reciprocal_SNR}
\end{figure}

In Fig.~\ref{MSE_reciprocal_SNR}, we plot the MSE performance of the derived semi-blind EM algorithm versus SNR for $L=8$ and $N=32$, where the SNR is defined as $10\log\frac{P_2}{\sigma^2}$. The channel estimates are obtained after $4$ iterations of the EM algorithm. For comparison, we also plot the MSE of the LS estimator that only uses the pilot samples. As we can see from Fig.~\ref{MSE_reciprocal_SNR}, the EM algorithm provides substantially higher accuracy than the LS estimator. Moreover, the gain in accuracy depends on the modulation order: the lower the modulation order the higher the gain.

\begin{figure}[!ht]
\centering
\includegraphics[width=3.4in, height=2.5in]{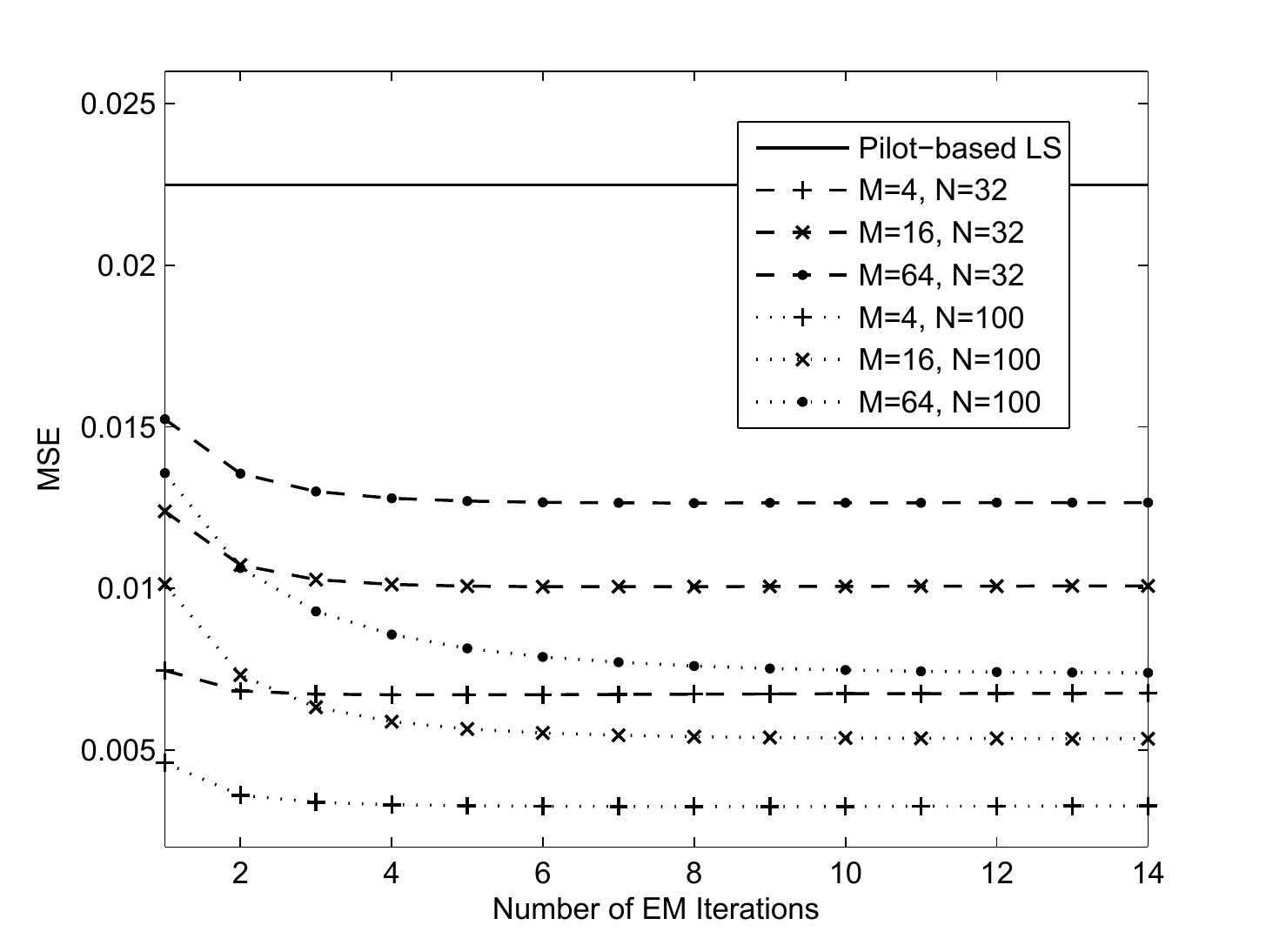}
\caption[MSE of EM algorithm vs. number of EM iterations (reciprocal channels, $15$dB).]{MSE performance of the EM algorithm for reciprocal channels plotted versus the number of EM iterations ($N=32, 100$, $L=8$, SNR $15$dB).}
\label{MSE_reciprocal_iterations}
\end{figure}

We finally consider the convergence of the EM algorithm. In Fig.~\ref{MSE_reciprocal_iterations}, we plot the MSE of the EM algorithm versus the number of iterations for $N=32$ and $N=100$, assuming 8 pilots and an SNR of $15$dB. Fig.~\ref{MSE_reciprocal_iterations} shows that the number of iterations needed for convergence is small for all modulation orders. In all cases, convergence is achieved within at most 12 iterations (as few as $4$ iterations are sufficient in some cases). Convergence becomes slightly slower as the modulation order increases and as the number of data samples increases. 

\section{Conclusions} 
\label{conclusions_ch6}

In this work, we derived the EM algorithm for semi-blind channel estimation in AF TWRNs assuming reciprocal flat-fading channel conditions. The resulting computational complexity of the EM steps is linear in the number of data samples. Using simulations, we showed that, even with a limited number of data symbols, the derived EM algorithm provides a substantial improvement in accuracy over the pilot-based LS estimator. Moreover, it requires only a small number of iterations to converge.



%


\bibliographystyle{IEEEtran}
\bibliography{IEEEabrv,MyReferences}

\end{document}